\documentclass[10pt, conference]{IEEEtran}

\usepackage[T1]{fontenc}
\usepackage[nolist]{acronym}
\usepackage{url}
\usepackage{booktabs}
\usepackage{color,soul}
\usepackage[table]{xcolor}
\usepackage{enumitem}
\usepackage{xr}
\usepackage{textcomp}
\usepackage{multirow}
\usepackage{dirtree}
\usepackage{url}
\usepackage{pdfpages}
\usepackage{graphicx}
\usepackage{subcaption}
\usepackage{booktabs}
\usepackage{siunitx} 
\usepackage{cleveref}
\usepackage[a4paper, left=1.6cm,right=1.7cm,top=1.8cm]{geometry}
\usepackage{listings}
\usepackage{color}

\definecolor{dkgreen}{rgb}{0,0.6,0}
\definecolor{gray}{rgb}{0.5,0.5,0.5}
\definecolor{mauve}{rgb}{0.58,0,0.82}

\newcommand{\takeaway}[1]{\textit{\textbf{Takeaway:}} \textit{#1}}

\usepackage{mdframed}
\definecolor{gray90}{gray}{0.90}
\newenvironment{summary}
{\noindent\begin{mdframed}[backgroundcolor = gray90, linecolor = black, innerleftmargin=1.5mm, innerrightmargin=1.5mm]}
{\end{mdframed}}

\begin{acronym}
  \acro{DNS}{Domain Name System}
\end{acronym}

\begin{document}

\title{\huge Through the Lens of Google CrUX: Dissecting Web Browsing Experience Across Devices and Countries}


\author{\normalsize\IEEEauthorblockN{Jayasree Sengupta\IEEEauthorrefmark{1},
Tanya Shreedhar\IEEEauthorrefmark{2},
Dinh Nguyen\IEEEauthorrefmark{3}, 
Robert Kramer\IEEEauthorrefmark{3}, and
Vaibhav Bajpai\IEEEauthorrefmark{4}}
\IEEEauthorrefmark{1} BIT Mesra, India;
\IEEEauthorrefmark{2}The University of Edinburgh, UK\\ 
\IEEEauthorrefmark{3}Technical University of Munich, Germany;
\IEEEauthorrefmark{4}Hasso-Plattner Institute, Germany
}


\maketitle
\IEEEpeerreviewmaketitle

\begin{abstract}

  User quality of experience in the context of Web browsing is being researched widely, with plenty of developments occurring alongside technological advances, not seldom driven by big industry players. With Google’s huge reach and infrastructure, the Chrome User Experience Report (CrUX) provides quantitative real-life measurement data of a vast magnitude. Analysis of this steadily expanding dataset aggregating different user experience metrics,  yields tangible insights into actual trends and developments. Hence, this paper is the first to study the CrUX dataset from the viewpoint of relevant metrics by quantitative evaluation of users’ Web browsing experience across three device types and nine European countries. Analysis of data segmented by connection type in the device dimension shows desktops outperforming other device types for all metrics. Similar analysis in the country dimension, shows North European countries (Sweden, Finland) having maximum 4G connections (85.99\%, 81.41\% respectively) and steadily performing 25\%-36\% better at the ${75}^{th}$ percentile across all metrics compared to the worst performing country. Such a high-level longitudinal analysis of real-life Web browsing experience provides an extensive base for future research.



\end{abstract}


\section{Introduction}\label{sec:introduction}
User Quality of Experience (QoE) can be described as "a subjective measure of customer's experiences" \cite{DBLP:conf/wwic/Martinez-YelmoSG10}. 
With the growth and development of the Internet, the desire for better understanding and metrication of the quality of experience now exists for many domains in the Internet ecosystem \cite{Moorsel}. 
%
User QoE for browsing the Internet is an actively researched field with particular interest from various big players such as Internet Service Providers (ISPs), Content Distribution Networks (CDNs), infrastructure providers, content originators like YouTube and Google, as well as the end-users themselves. With web browsing being one of the most dominant activities on cellular networks \cite{DBLP:conf/mobicom/BalachandranAHPSVY14}, accompanied by an increase of mobile web data usage over the years \cite{meeker19}\cite{itu}\cite{ciscoWP} and mobile phones steadily having the highest usage densities worldwide \cite{statCounter}, research highly benefits from measuring the user quality of experience. 

In this context, the Google CrUX dataset is a vastly large public dataset that uses Google’s large-scale Internet measurement architecture~\cite{paxson} to aggregate browsing data and derive invaluable real-life user quality of experience metrics across all device types.
Yet, the CrUX report has limited scientific work \cite{Ruth_WWW_IMC'22}\cite{Ruth_website_IMC'22} done on it to date (see: §\ref{sec:relatedwork}). Additionally, 
the raw metrics, such as First Paint (FP), DOM Content Loaded (DCL), etc. gathered through browser, i.e. Google Chrome usage is yet to be explored in depth. It can be put into applicable contexts and paired with the dataset’s ongoing aggregation over time, capturing longitudinal developments. This makes for a powerful resource 
as it will provide important insights to the site owners for further development, especially as there is an emerging interest in in-the-wild measurements \cite{DBLP:conf/hotnets/BalachandranSASSZ12}\cite{DBLP:journals/ccr/BustamanteCF17}\cite{DBLP:journals/cm/IckinWFJHD12}\cite{DBLP:conf/sigcomm/KhokharSSB17}. 

\begin{figure}[!t]
    \centering
    \begin{subfigure}[t]{0.48\textwidth}
        \centering     \includegraphics[width=0.97\columnwidth]{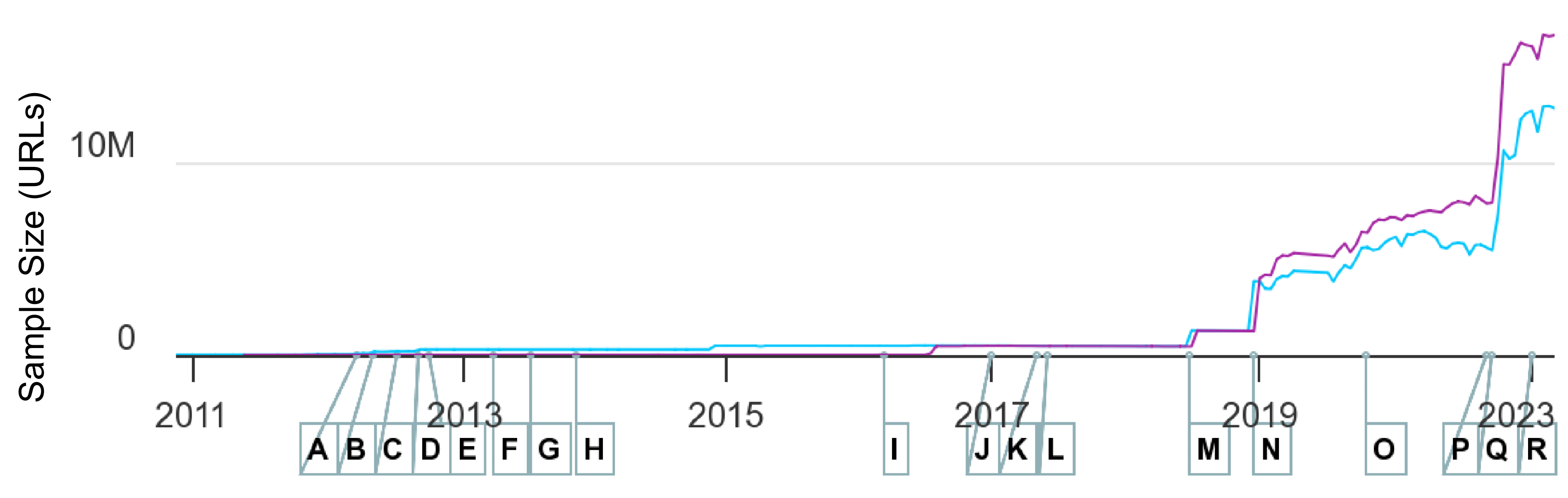}
        \caption{Timeseries of Sample Size}
    \end{subfigure}%
    \vspace{1em}
    \begin{subfigure}[t]{0.48\textwidth}
        \centering      \includegraphics[width=0.97\columnwidth]{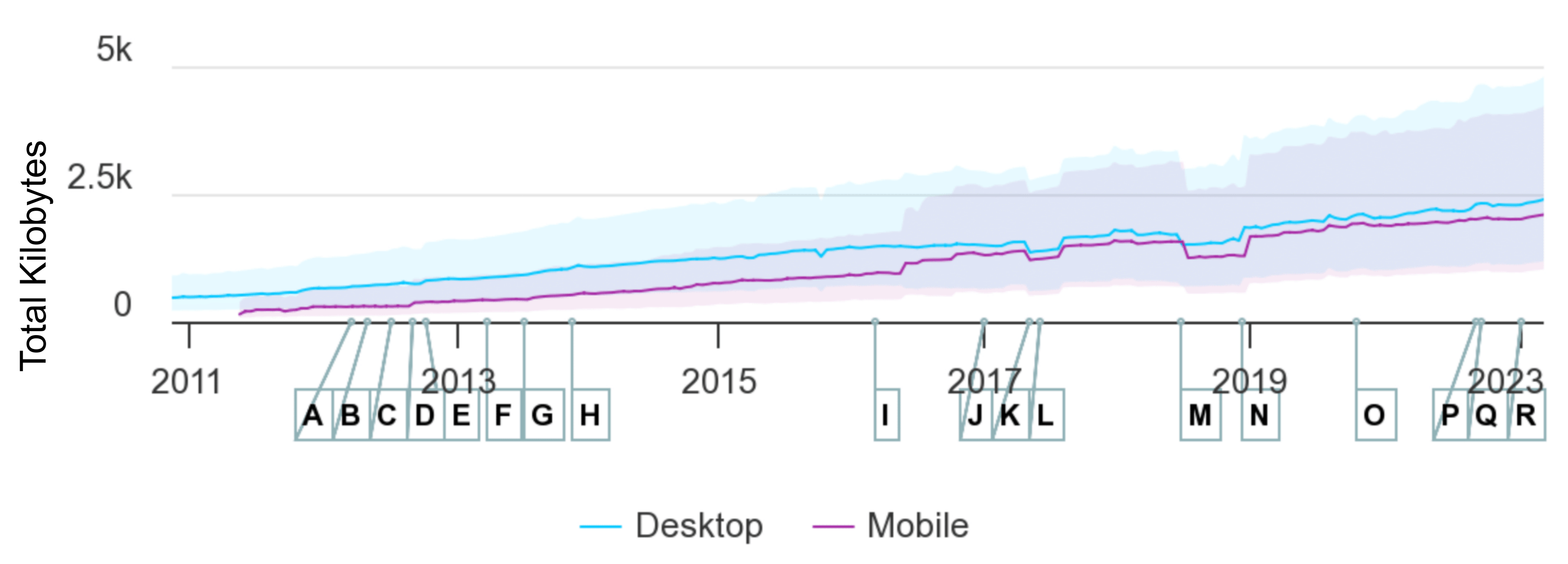}
        \caption{Timeseries of Total Kilobytes}
    \end{subfigure}
    \caption{\small Growth of the Internet from 2010 to 2023 both in terms of number of URLs (a, top) and median website size in KB (b, bottom) as detected by HTTParchive \cite{HTMLarchive2}.}
    \label{fig:html}
\end{figure}

\begin{figure*}[!t]
    \centering
    \vspace{2mm}
    \includegraphics[width=\textwidth]{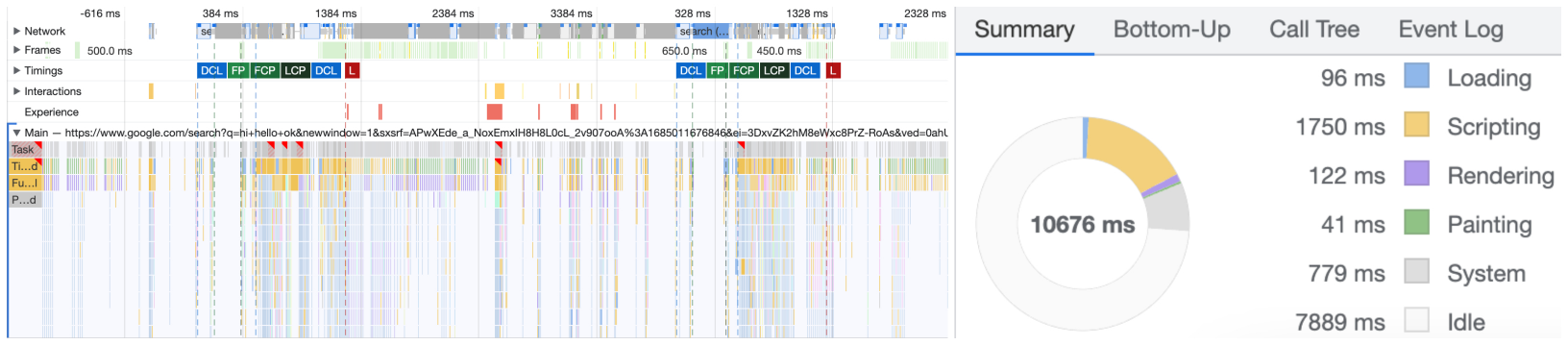}
    \caption{\small Performance graph generated by Google Chrome's built-in developer tools for https://www.google.com load timings, showcasing its relative temporal relation on a 4G type connection across different metrics.}
    \label{fig:net-graph-04-06-2019}
\end{figure*}



The Internet has grown tremendously over the years (see Fig. \ref{fig:html}), especially in the last decade (from 2018 onwards) with an exponential increase in URLs.
With the popularity of smartphones, mobile Internet has also seen a steady development and received widespread user acceptance as already predicted in \cite{DBLP:journals/dss/KimCG07}. The explosive growth of the mobile Internet, followed by a linear increase in tablet ownership further reinforced the mobile Internet user base. The Internet has now reached a penetration of over 3.8 billion people, the first time numerically over 50\% of the world’s population is using the Internet according to Meeker \cite{meeker19} which has seen a steady increase from 24\% in 2009. According to independent data from StatCounter \cite{statCounter}, mobile Internet and desktop Internet users have a similar share at 48\% and 49\% respectively, showcasing the dichotomy in the current Internet device landscape. Thus, efforts (see: §\ref{sec:relatedwork}) have been made to factor users in more real-life data measurements such as those conducted by Ickin et al. \cite{DBLP:journals/cm/IckinWFJHD12} for mobile application quality of experience. Also, efforts to improve user experience in mobile Internet browsing regarding loading times and latency have been reviewed by Cazanas et al. \cite{cazanas2017strategies}, identifying major web design strategies aimed at heightening mobile user quality of experience. However, such a comprehensive study on users browsing experience (in Google Chrome) across different metrics and dimensions over the CrUX dataset has not been conducted yet, which serves as the main motivation for this paper.






This paper utilizes real-life user measurement data aggregated in the Google CrUX dataset and highlights various trends and developments, providing a longitudinal overview of user quality of experience to enable future research.  Furthermore, analysis near the top level of the data among given dimensions showcases different aspects and areas of research where the CrUX dataset will be useful. This is achieved by analyzing all metrics in the dataset segmented along with all dimensions. In addition, we further reaffirm existing observations and developments in research. In summary, we make the following key contributions in this paper:
 \begin{itemize}[leftmargin=*]
     \item \textbf{Measurement:} The convoluted methods and options of acquiring, aggregating and segmenting the CrUX dataset are explored (see: \S\ref{sec:methodology}) with a focus on reproducibility. The dataset is segmented by effective connection type to study development of user experience across three device types and nine European countries. We are the first to conduct a detailed metric-wise analysis and compare its performance in these dimensions. This yields tangible insights into users' actual Web browsing experiences under different scenarios.
     \item \textbf{Findings:} We show (see: \S\ref{sec:analysis}) that despite phone density (64.8\%) being higher than desktops (31.5\%) in the CrUX dataset, the measurement across metrics reveals that desktop devices outperform other device types. We also observe that North European countries, i.e. Sweden and Finland have maximum 4G connections (85.99\% and 81.41\%, respectively). At the ${75}^{th}$ percentile, these two countries perform 25\%-36\% better than the worst performing country Italy, across all metrics. These findings reflect new insights into the device usage patterns and user experience along with the leading countries and geographical locations driving such patterns. To support reproducibility, we will release the measurement setup and analysis scripts publicly on GitHub.
     
 \end{itemize}

\section{Methodology}\label{sec:methodology}
The \textbf{Chrome User Experience Report} (CrUX Report) aggregates real-life Internet usage quality of experience data collected from Google Chrome users who opted to send their Google usage statistics \cite{GCRUX}. While the usage statistics encompass data about system information, preferences, user interface feature usage, responsiveness, performance, and memory usage \cite{GPRIV}, the CrUX report enlists only a subset or derivative of such data.
Google exposes the data via \textit{PageSpeed Insights} providing URL-level data derived from CrUX via web frontend or API \cite{psiGoogle}, and \textit{BigQuery} \cite{bigQueryDoc}, providing the whole dataset as well as an SQL-like interface to perform queries on the dataset \cite{bigQuery}. On the highest level, the data is either available as a globally aggregated set called \textit{all}, or segmented by countries.
In either dataset, there are tables for each month in which the actual metrics and dimensions are included, aggregated by \textit{origins}. 



\subsection{Metrics}
The metrics available in the CrUX dataset \cite{GCRUX} mainly load performance-focused data exclusively on a time-based measurement. Due to the data collection process, metrics are comprised of key timings in the \textit{Critical Rendering Path} 
and enable deeper insight and optimization in website engineering (see: Fig. \ref{fig:net-graph-04-06-2019}). 
Each metric is represented as a histogram nested into a structure yielding three nodes having:
\begin{itemize}[leftmargin=*]
\item \textbf{start}: Lower time boundary of the bin in milliseconds. Data: Integer value
\item \textbf{end}: Upper time boundary of the bin in milliseconds. Data: Integer value
\item \textbf{density}: Normalized density in the current bin. Data: Floating point number between 0 to 1
\end{itemize}

\begin{figure*}[t!]
  \centering
  \begin{subfigure}[b]{0.48\linewidth}
    \includegraphics[width=0.95\linewidth]{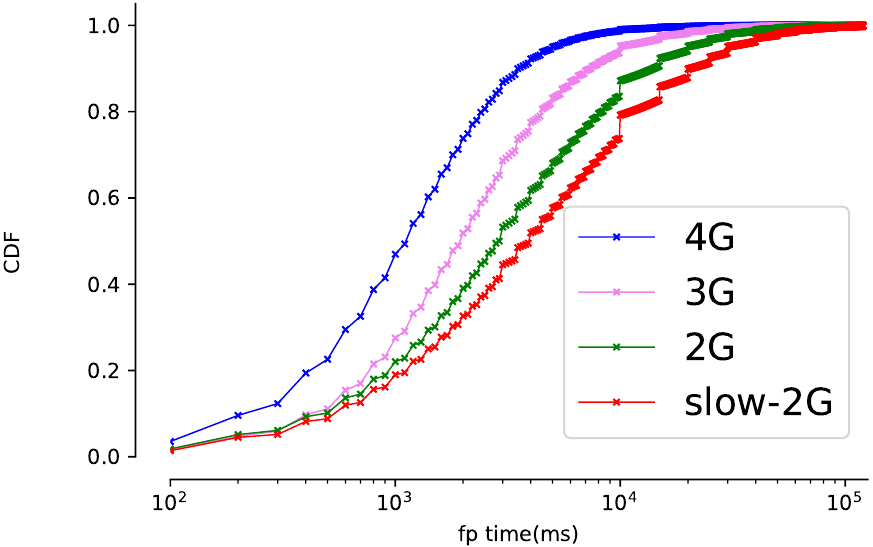}
    \caption{\sl \small First Paint (FP)}
  \end{subfigure}
  \begin{subfigure}[b]{0.48\linewidth}
    \includegraphics[width=0.95\linewidth]{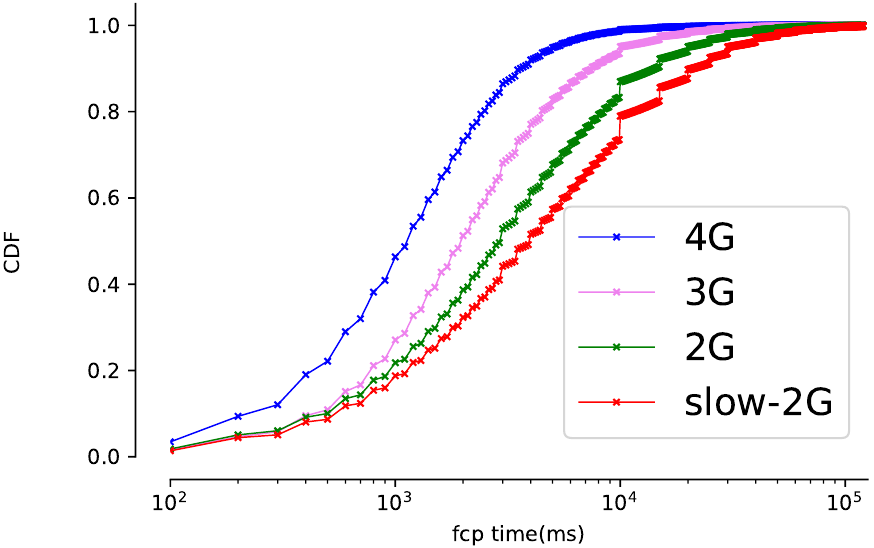}
    \caption{\sl \small First Contentful Paint (FCP)}
  \end{subfigure}
  \begin{subfigure}[b]{0.48\linewidth}
    \includegraphics[width=0.95\linewidth]{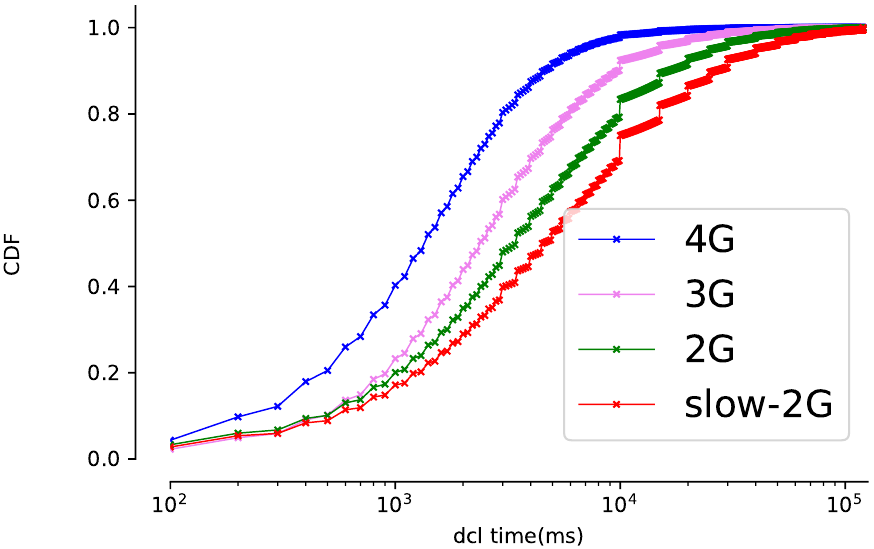}
    \caption{\sl \small DOM Content Loaded (DCL)}
  \end{subfigure}
  \begin{subfigure}[b]{0.48\linewidth}
    \includegraphics[width=0.95\linewidth]{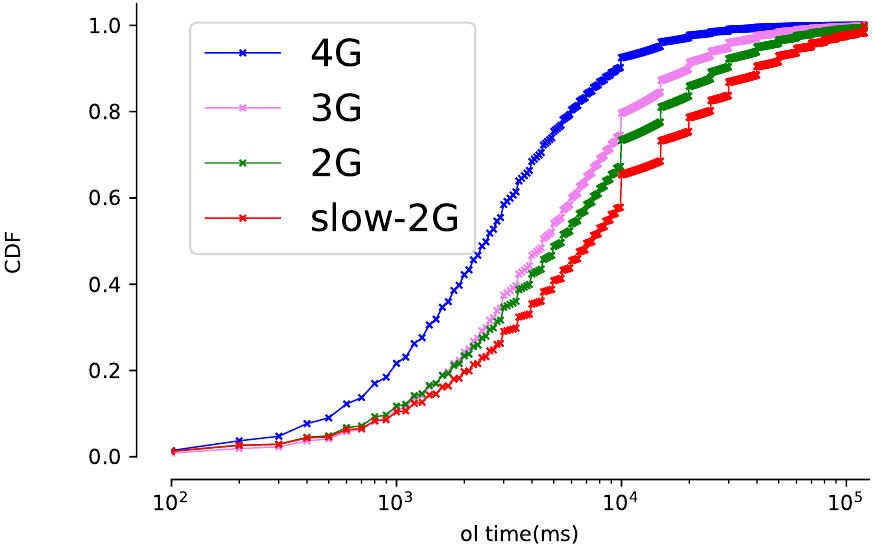}
    \caption{\sl \small Onload (OL)}
  \end{subfigure}
  \caption[CDF of all metrics among connection speed types]{\small Metrics across connection types for the dataset. Higher speeds result in direct improvement for every metric.}
  \label{fig:metrics-among-speeds}
  \vspace{-3mm}
\end{figure*}


\subsubsection{First Paint (FP) \cite{W3CAPI}}
Describes the point in time when anything visually discerning (to an empty page) is rendered on the screen
. This, together with \textit{first contentful paint} generally marks the quality of experience notion of first visual feedback, notifying the user that "something is happening" 
and is thus a measurable property for delay in user perception. 

\subsubsection{First Contentful Paint (FCP)}
 Refers to the point in time when the first content from the \textit{Document Object Model} (DOM) is rendered. Both \textit{FP} and \textit{FCP} impact traditional quality of experience notions of user-perceived latency \cite{DBLP:journals/internet/Mao16}
and start-up time \cite{DBLP:conf/wwic/KuipersKVB10}, 
affecting quality of experience directly \cite{DBLP:journals/comsur/BriscoeBPHRTGFG16}. Particularly more than other metrics, \textit{FCP} correlates to the delay users tolerate in general web usage \cite{DBLP:journals/jais/GallettaHMP04}.

\subsubsection{DOMContentLoaded \cite{HTMLS}}
Exact time when the complete HTML file has been parsed, regardless of other ongoing loading processes. After this event finishes, the render tree can be built. Websites achieve better speed and quality of experience by optimizing prior DCL events \cite{dclVarvy}. 
For engineering purposes, minimising \textit{DCL} timing results in earlier \textit{onload} timings due to faster rendering. While both metrics either directly or indirectly affect the overall load time of a website, \textit{DOM content loaded} timings are improved solely by website engineering \cite{domMozilla}.

\subsubsection{Onload (OL)}
Specifies the point in time when the whole page including all dependent resources have been loaded, both in terms of content as well as visibly. 
This plays well into quality of experience metrics such as successful download completion probabilities and availability of service \cite{Moorsel} aside from being a self-defined \textit{load time} metric for websites, capturing the impact of bandwidth \cite{DBLP:journals/ccr/Ford14}.


\subsection{Dimensions}
Dimensions provide useful segmentation possibilities of the metrics.

\noindent\textbf{Effective Connection Type:}  
It is not necessarily the actual connection type, but classified by down speeds and round-trip-times. This means a data entry marked as "3G" could be a slow desktop fixed connection despite all connotations, while a data entry marked as "2G" could be a high bandwidth connection with unusually high round-trip-times. As defined by the W3C Network API \cite{netAPI}. "Offline" and "slow-2G" will be considered fringe cases in this paper. While "slow-2G" limits itself to a very slow down speed bandwidth of 50 kbps and round-trip-time of 2000 ms, "offline" includes Progressive Web Application's support for offline usage \cite{hume} which was deemed unrepresentative for current work

\noindent\textbf{Device Type:}
Classified via User-Agent string \cite{RFCua}
from the Chrome browser \cite{ChromeUA}. 
In the dataset, this dimension is simply injected into three possible string values: \textit{"desktop"}, \textit{"phone"} and \textit{"tablet"}.

\begin{figure*}[!t]
  \centering
  \begin{subfigure}[b]{0.48\linewidth}
    \includegraphics[width=\linewidth]{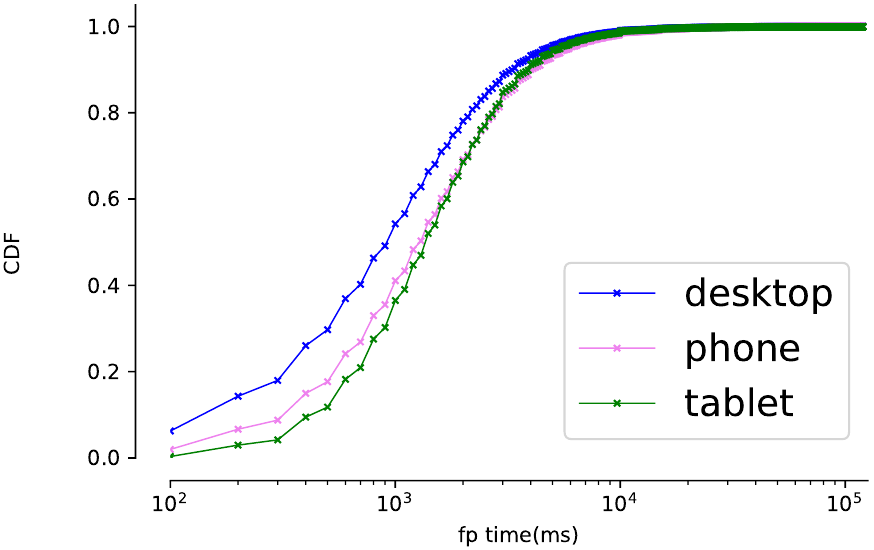}
    \caption{First Paint (FP)}
  \end{subfigure}
  \begin{subfigure}[b]{0.48\linewidth}
    \includegraphics[width=\linewidth]{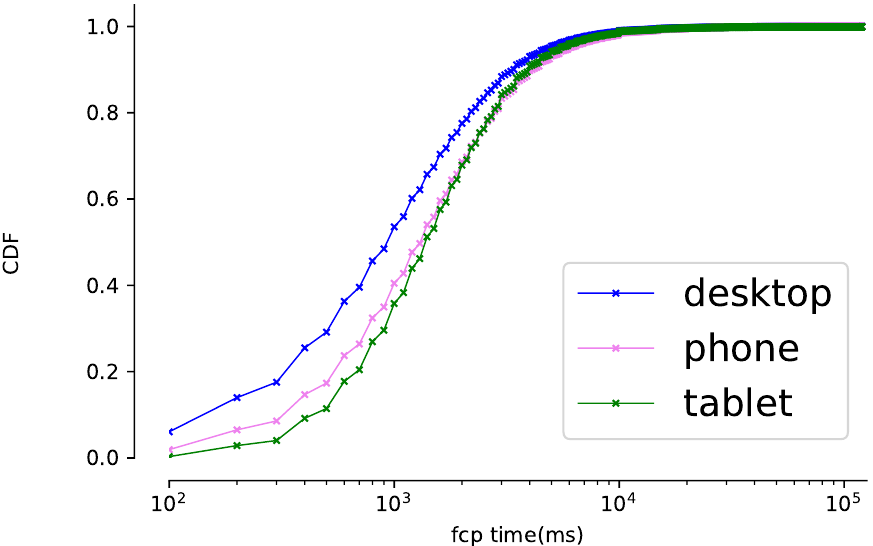}
    \caption{First Contentful Paint (FCP)}
  \end{subfigure}
  \begin{subfigure}[b]{0.48\linewidth}
    \includegraphics[width=\linewidth]{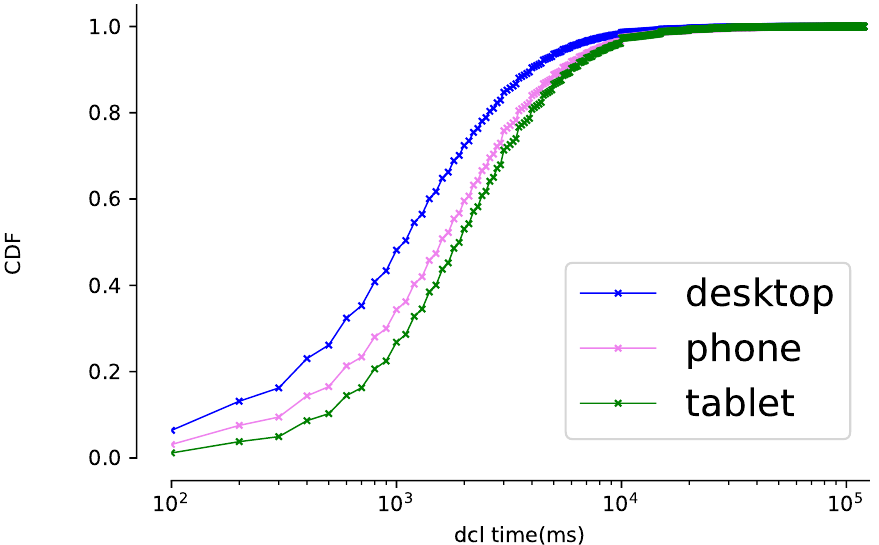}
    \caption{DOM Content Loaded (DCL)}
  \end{subfigure}
  \begin{subfigure}[b]{0.48\linewidth}
    \includegraphics[width=\linewidth]{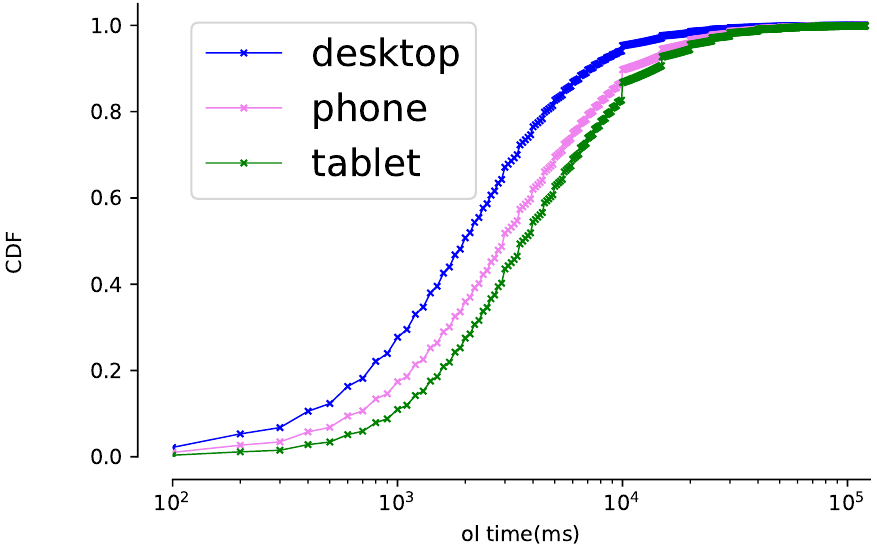}
    \caption{Onload (OL)}
  \end{subfigure}
  \caption[CDF of all metrics among devices]{\small CDFs of all metrics across devices types, for the whole dataset duration. A clear classification is visible - desktop outperforms phones, which in turn outperform tablets - for all metrics.}
  \label{fig:metrics-among-devices}
  \vspace{-3mm}
\end{figure*}

\noindent\textbf{Country:}
Dependent on the IP address of the device where the content is being delivered to. Notably, this dimension is not modeled inside of the tables but rather achieved through aggregation of country-specific sub-datasets of aforementioned metrics. 



\noindent\textbf{Origin:} An origin is a website, as known to Google's web crawlers. Metrics are aggregated by origin where the data is a string containing web address of the site.

\noindent\textbf{Timeframe:} CrUX data can be dissected into time spans of whole months by aligning with the Gregorian Calendar. 
However, the time frame is not embedded into the tables, it is rather aggregated by months at the lowest extra-tabular level. 

\subsection{Distribution of Connection Types}\label{sec:metcon}

There is an increase of 4G density with time and network infrastructure improvement both for landline \cite{DBLP:journals/dss/KimCG07} 
and mobile Internet 
\cite{itu}\cite{DBLP:journals/jcmc/HumphreysPK13}\cite{DBLP:journals/cm/FettweisA14}.
Owing to the high variance in bandwidth between the connection types, segmentation by connection types provides useful insight. The metrics exhibit a predictable pattern as in faster speeds provide faster timings across all metrics (see Fig. \ref{fig:metrics-among-speeds}), which translates directly to more favourable quality of experience, showcasing a correlation; a finding congruent with existing notions of higher bandwidths, thereby improving quality of experience by reducing user cancellation rates. 
It is stressed that this pattern might implicitly result from the effective connection type classification not only by bandwidth but also by round-trip-time \cite{netAPI}. 
When comparing the betterment of the metrics along the different connection types, a slight rate decrease can be observed for all metrics while the \textit{onload} metric continues improving at a better rate than all other metrics, reaffirming the implication that the \textit{onload} timing is more bandwidth-dependent than the other metrics, 
not to mention subject to quality of service related implications like jitters or packet loss\cite{DBLP:journals/network/FiedlerHT10}\cite{DBLP:conf/wwic/KuipersKVB10}\cite{DBLP:conf/wwic/Martinez-YelmoSG10}. 
Given the steady increase of 4G data compared to stagnating data of all other connection types, 
this also means that general research will carry most meaning for 4G type connections while the other speeds are special or fringe case analyses.

\noindent\textbf{First Paint:}
Segmentation along the speed dimension shows a roughly steady, slightly downwards trend towards the $1000\ ms$ mark for 4G connections. 
The steady FP timing aligns with the notion of FP being infeasible much earlier due to unsolved network latency constraints caused by adding bandwidth \cite{DBLP:journals/comsur/BriscoeBPHRTGFG16}. 
Analysis of 3G as well as 2G speeds showcase an increase in average FP time, suggesting website and content distribution development following the general Internet speed trends. 



\begin{figure*}[!t]
\vspace{10mm}
 \centering
 \begin{subfigure}[b]{0.32\linewidth}
   \includegraphics[width=\linewidth]{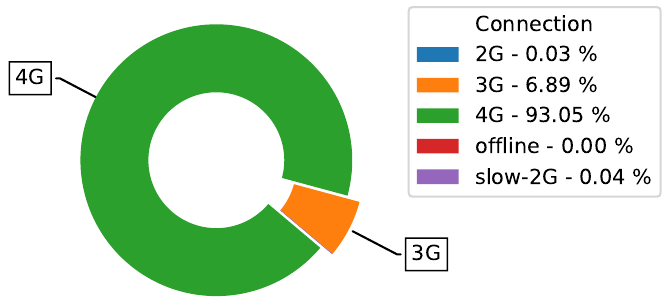}
   \caption{Phone}
 \end{subfigure}
 \begin{subfigure}[b]{0.32\linewidth}
   \includegraphics[width=\linewidth]{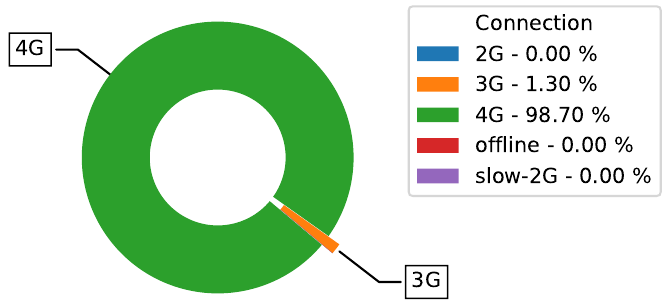}
   \caption{Desktop}
 \end{subfigure}
 \begin{subfigure}[b]{0.32\linewidth}
   \includegraphics[width=\linewidth]{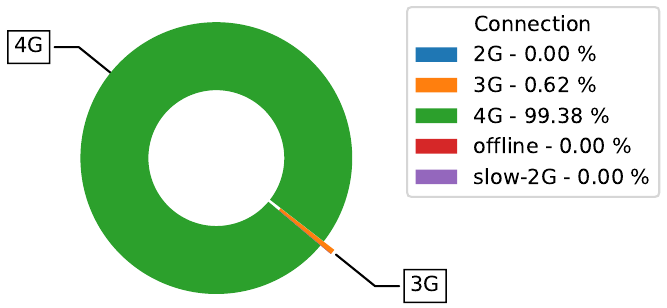}
   \caption{Tablet}
 \end{subfigure}

 \caption[Piecharts of connection type proportions]{\small Proportions of connection speed type segmented by device. 
 A common trend among all devices is the increasing predominance of 4G type connections.}
 \label{fig:PC-Conn-for-dev}
\end{figure*}

\noindent\textbf{First Contentful Paint:}
\textit{FCP} timings in the CrUX dataset behave similarly to the \textit{FP} ones.
Thus, for connection type segmentation, \textit{FCP} and \textit{FP} metrics can be used for the same analyses.

\noindent\textbf{DOM Content Loaded:}
This metric is more important for front-end engineers, as the timing is often waited for by Javascript frameworks before executing scripts \cite{crpGoogle}. 
As the DOM-tree's complexity impacts (browser) computing and secondary effects more so than it effects network bandwidth \cite{exDomGoogle}, the timings are mostly steady albeit at different scales for 4G and 3G. 
For 2G connection types, which are defined to have a \textit{maximum} downspeed of 70 kbit/s, average DOM-tree sizes seem to impact the timing, showing a slow increase. 


\noindent\textbf{Onload:}
This metric shows slight improvement over time for 4G connections. 
Expectantly, 3G data shows about constant rates up to a slight decrease in onload performance near the end of the observed timeframe, while 2G speeds showcase worse performance over time, reaffirming speeds being a deciding factor for the onload metric \cite{DBLP:journals/comsur/BriscoeBPHRTGFG16}. 


\section{Analysis}\label{sec:analysis}

For analysis of temporal data, the dataset is segmented along the date dimension which aggregates data by month.
The CrUX real-life data behaves as expected considering definitions (see Section \ref{sec:methodology}) and assumptions regarding the timing metrics, with the order of timings being \textit{first paint}, \textit{first contentful paint}, \textit{DOM content loaded} and \textit{onload}. The \textit{first paint} timing is usually very close to the \textit{first contentful paint} timing. As an example, the medians are less than one bin size (less than 100 ms) apart. 
As the \textit{first contentful paint} timing is considered as meaningful as the \textit{first paint} timing \cite{gtMet}
, interpretation will be considered applicable on both; additionally, both generally correlate with established notions of delivery time in a quality of experience context. 

\subsection{Metrics Across Device Types}\label{sec:metdev}

The devices dimension has a clear three value classification with no semi-open interval ranges.

\subsubsection{General Analysis}
Data for tablets in the CrUX dataset make up less than 1\% at all time samples. 
Correspondingly, desktop density averages to 35.1\% leaving phone density at about 64.8\%. 
Looking at all metrics segmented by devices, desktop devices display a distinct lead over other device types in any metric (see Fig.  \ref{fig:metrics-among-devices}), which can be assumed to correlate with generally higher network speeds (see §~\ref{sec:metcon}) usually used on a desktop via landline Internet connections. 
As phones are known to be used with mobile Internet more than desktop and tablet devices are \cite{itu}, there are quality of service aspects \cite{DBLP:journals/network/FiedlerHT10}\cite{DBLP:conf/wwic/KuipersKVB10} 
to be considered as a factor responsible for slower timings.
Although tablet devices have an even higher 4G usage density than desktop devices, they have the slowest timings among all devices, hence, requiring further investigation.

\subsubsection{Temporal Analysis}\label{metdevtemp}
Over the analyzed time period, 
tablet usage is always below 1\% when compared to phone and desktop usage, indicating a strong and steady preference of devices towards phones in the area of mobile Internet browsing usage. When looking at phone-to-desktop-ratio, there is no clear trend visible but 
desktop devices never amounted to more than 46.67\% of devices used to browse the web via Google Chrome and averages to 35.09\% over the complete timespan, making phones the most prevalent device as has been previously predicted \cite{DBLP:journals/jcmc/HumphreysPK13}\cite{meeker10} 
and observed \cite{itu}\cite{meeker19} 
in research. In line with observations made in §~\ref{sec:metcon}, the density of 4G type connections is ever rising for each device type segmented by month. As expected, availability of 4G speeds on mobile Internet lags behind that of desktop devices (see Fig. \ref{fig:PC-Conn-for-dev}) \cite{itu}, especially due to the latter's usual connection via landline.




\noindent\textbf{First Paint:}
The median of the first paint timings, 
remain fairly stead within 1000ms to 2000ms range for all device types, with desktop being slightly faster than its mobile counterparts. This supports the findings in §~\ref{sec:metcon}, given the prevalence of 4G connection types (see Fig. \ref{fig:PC-Conn-for-dev}), being mainly limited by a baseline network latency \cite{DBLP:journals/comsur/BriscoeBPHRTGFG16} 
rather than the device type. 


\noindent\textbf{First Contentful Paint:}
First contentful paint analysis yields, again, similar results to first paint analysis according to the similar nature of the timings. 
Steady timings over the entire timeframe for each device type is expected considering consolidation of connection speeds. Median timings near the 1000ms mark coincide with general notions of short delays or fast response times in Internet browsing \cite{walton}\cite{DBLP:journals/jais/GallettaHMP04}.

\begin{table*}[!t]
\centering
\vspace{5mm}
\caption{\small Country-wise connection distribution in percent. Sweden and Finland have the highest (shown in blue) 4G connection requests, whereas France and Italy having the lowest (shown in grey) percentage of requests.}
\label{tab:CD_country}
\resizebox{0.90\textwidth}{!}{%
\begin{tabular}{|c|ccccccccc|}
\hline
 & \multicolumn{9}{c|}{\textbf{Countries (\%)}} \\ \cline{2-10} 
\multirow{-2}{*}{\textbf{\begin{tabular}[c]{@{}c@{}}Connection\\ Types\end{tabular}}} & \multicolumn{1}{c|}{\textbf{Germany}} & \multicolumn{1}{c|}{\textbf{Great Britain}} & \multicolumn{1}{c|}{\textbf{France}} & \multicolumn{1}{c|}{\textbf{Italy}} & \multicolumn{1}{c|}{\textbf{Spain}} & \multicolumn{1}{c|}{\textbf{Sweden}} & \multicolumn{1}{c|}{\textbf{Finland}} & \multicolumn{1}{c|}{\textbf{Poland}} & \textbf{Romania} \\ \hline
4G & \multicolumn{1}{c|}{78.4} & \multicolumn{1}{c|}{78.44} & \multicolumn{1}{c|}{\cellcolor[HTML]{9B9B9B}70.25} & \multicolumn{1}{c|}{\cellcolor[HTML]{C0C0C0}69.7} & \multicolumn{1}{c|}{76.52} & \multicolumn{1}{c|}{\cellcolor[HTML]{38FFF8}85.99} & \multicolumn{1}{c|}{\cellcolor[HTML]{96FFFB}81.41} & \multicolumn{1}{c|}{73.86} & 75.06 \\ \hline
3G & \multicolumn{1}{c|}{21.12} & \multicolumn{1}{c|}{21.01} & \multicolumn{1}{c|}{29.0} & \multicolumn{1}{c|}{29.29} & \multicolumn{1}{c|}{23.33} & \multicolumn{1}{c|}{13.96} & \multicolumn{1}{c|}{18.57} & \multicolumn{1}{c|}{25.56} & 24.5 \\ \hline
2G & \multicolumn{1}{c|}{0.19} & \multicolumn{1}{c|}{0.3} & \multicolumn{1}{c|}{0.54} & \multicolumn{1}{c|}{0.85} & \multicolumn{1}{c|}{0.1} & \multicolumn{1}{c|}{0.02} & \multicolumn{1}{c|}{0.01} & \multicolumn{1}{c|}{0.38} & 0.2 \\ \hline
slow-2G & \multicolumn{1}{c|}{0.29} & \multicolumn{1}{c|}{0.24} & \multicolumn{1}{c|}{0.21} & \multicolumn{1}{c|}{0.17} & \multicolumn{1}{c|}{0.04} & \multicolumn{1}{c|}{0.03} & \multicolumn{1}{c|}{0.01} & \multicolumn{1}{c|}{0.2} & 0.24 \\ \hline
offline & \multicolumn{1}{c|}{0.02} & \multicolumn{1}{c|}{0.0} & \multicolumn{1}{c|}{0.01} & \multicolumn{1}{c|}{0.0} & \multicolumn{1}{c|}{0.0} & \multicolumn{1}{c|}{0.0} & \multicolumn{1}{c|}{0.0} & \multicolumn{1}{c|}{0.0} & 0.01 \\ \hline
\end{tabular}%
}
\end{table*}


\noindent\textbf{DOM Content Loaded:}
DCL  timing is dependent on network speed, latency and especially structure of the webpage
, suggesting most devices not being a bottleneck in hardware aspects. As this segmentation aggregates without considering connection speed types, barely any change can be seen from the start of the non-mass-aggregation part of the data.
This observation, among all devices, is congruent to the premise of median website data fetched barely changing (increase in 12.5\% for mobile and 9.1\% for desktop) in the observed timeframe \cite{HTMLarchive}.


\noindent\textbf{Onload:}
The onload metric 
shows varying degrees of improvements over time. Mobile devices (phone and tablet) are seen to have visible improvements, going from ~3200 to ~2700 and ~5800 to ~3500 respectively. 
Interestingly enough, desktop devices have a very flat downwards curve, 
finishing most onloads in the median ~2400 range.
This suggests several possible hypotheses:
\begin{itemize}[leftmargin=*]
\item Desktop PCs are usually connected via fast 
and stable fixed Internet connections, thus providing best OL performances with least variance (among devices). 
\item Tablet devices predominantly use home WIFI connections usually connected via the same fast and stable landline Internet connections too, they differ to desktop PCs in two aspects: machine power and general defaulting to mobile versions of websites.
\item Thus, the notable bigger improvements of mobile device OL performance could indicate optimization or even just emerging availability of mobile website versions \cite{cazanas2017strategies} or responsive design advancements. 
\item Mobile Internet availability and speeds are improving. 
\end{itemize}

\begin{summary}
\takeaway{Phone density is highest at around $64.8\%$, followed by desktop density at $31.5\%$. Looking at all metrics segmented by devices, desktop devices display a distinct lead over other types. Although tablet devices have an even higher 4G usage density than desktop devices, they have the slowest timings amongst all.}
\end{summary}

\subsection{Metrics across different European countries}

This section analyses how different European countries compete with each other in terms of mobile QoE. For the analysis following countries were chosen: Germany (DE), Great Britain (GB), France (FR), Spain (ES), Italy (IT), Sweden (SE), Finland (FI), Poland (PL), Romania (RO). 


\subsubsection{General Analysis} The distribution of different connection types across countries within the EU is shown in Table  \ref{tab:CD_country}. Looking at the data, it is evident that only a very small percentile comes from connections, which are slower than `3G’ namely `offline’, `slow-2G’ or `2G’. It is observed that most connections of these types combined are found in Italy with 1.02\% and France with 0.75\%. Great Britain, Germany, Romania and Poland are in the middle with 0.54\%, 0.5\%, 0.45\% and 0.4\% respectively. Spain has fewer requests than the middle of the pack with 0.14\%, but the two countries with the lowest amount of connections slower than `3G’ are in Sweden and Finland with 0.05\% and 0.02\% by quite a bit. As majority of the requests come from the connection types `3G’ and `4G’ they correlate with each other. Just as with the slower connection types, Sweden and Finland are also ahead here with 85.99\% `4G’ and 13.96\% `3G’ requests and 81.41\% `4G’ and 18.57\% `3G’ requests respectively. Great Britain and Germany follow after that with 78.44\% `4G’ and 21.01\% `3G’ requests and 78.4\% `4G’ and 21.12\% `3G’ requests respectively. Spain, Romania and Poland are the next contenders having 76.52\% `4G’ and 23.33\% `3G’ requests, 75.06\% `4G’ and 24.5\% `3G’ requests and 73.86\% `4G’ and 25.56\% `3G’ requests correspondingly. Again, France and Italy are last in the list with the lowest amount of `3G’ and `4G’ requests, France with 70.25\% `4G’ and 29.0\% `3G’ requests and Italy with 69.7\% `4G’ and 29.29\% `3G’ requests. Overall, the countries differ a lot when it comes to the distribution of the different connection types.

\subsubsection{Temporal Analysis} As observed from the above analysis, the amount of `3G' connections is too low to be compared, hence in this section, the measurement is restricted to `4G' only. 

\begin{figure}[!t]
    \centering
    \begin{subfigure}[t]{0.48\textwidth}
        \centering
        \includegraphics[width=\textwidth]{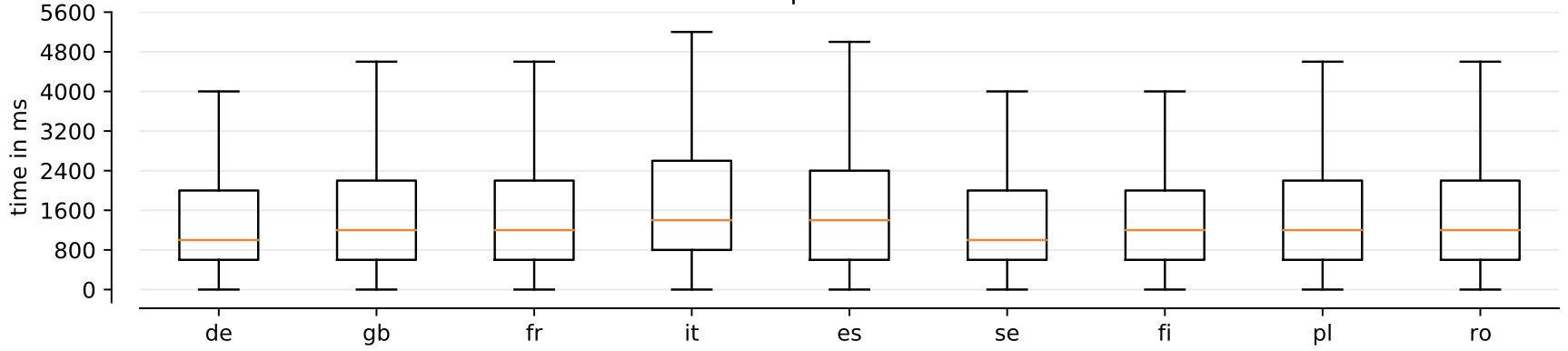}
        \caption{First Contentful Paint}
        \label{fig:Boxplot-FCP-country}
    \end{subfigure}%
    \vspace{1em}
    \begin{subfigure}[t]{0.48\textwidth}
        \centering
        \includegraphics[width=\textwidth]{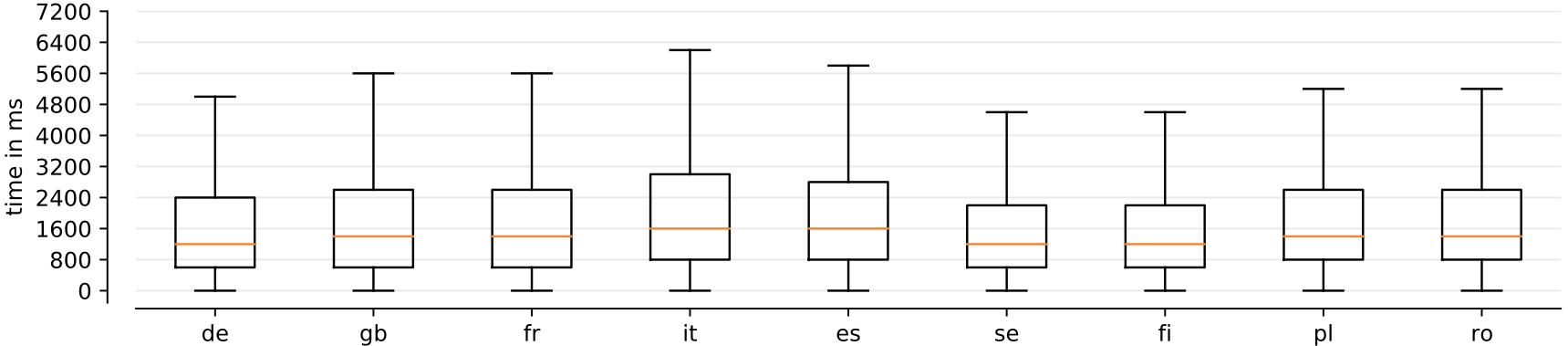}
        \caption{DOMContentLoaded}
        \label{fig:Boxplot-DCL-country}
    \end{subfigure}
    \vspace{1em}
    \begin{subfigure}[t]{0.48\textwidth}
        \centering
        \includegraphics[width=\textwidth]{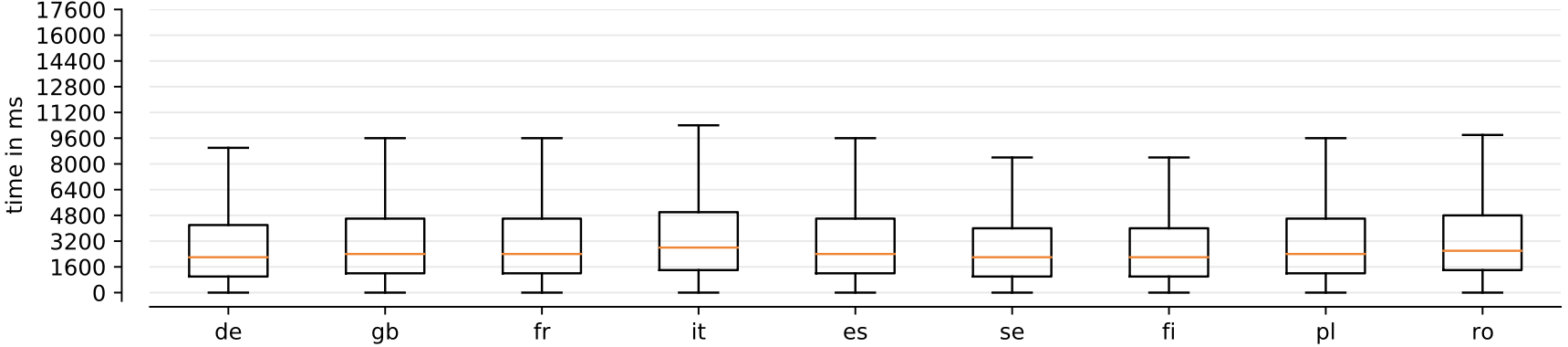}
        \caption{Onload}
        \label{fig:Boxplot-OL-country}
    \end{subfigure}
    \caption{\small Country-wise distribution of different metrics for 4G connection type. Sweden and Finalnd are the best performing countries providing good user experience while Italy is the worst.}
    \vspace{-1em}
\end{figure}

\noindent\textbf{First Contentful Paint:} Fig. \ref{fig:Boxplot-FCP-country} shows the normalized distribution where all countries share the same minimum at 0ms. The ${25}^{th}$ percentile is fixed at 600ms for all countries except Italy which is 800ms. Germany and Sweden have a median at 1000 ms, whereas others are worse by 200ms, except Italy and Spain which records the highest median values at 1400 ms. When looking at the ${75}^{th}$ percentile the graphs start to differ more with Germany, Sweden and Finland at 2000ms whereas Italy at 2600 ms. The same picture is reflected for the maximum with the same three countries on top of the standings at 4000 ms, whereas  Italy has the worst connections with 5200ms. All other countries have metrics in between these values with Spain performing worse compared to others.

For this metric, Google proposes to interpret the key indicators at the ${75}^{th}$ percentile as
\cite{WebVitalsGoogle}
: loading the paint can be fast (< 2 seconds), moderate (2-4 seconds) and slow (> 4 seconds). Thus, only Sweden, Finland and Germany have a `good’ user experience (i.e. fast FCP loading) while barely managing to reach the goal. All other countries lack a bit behind and might need improvement. However, it is important to mention that even the worst country, Italy, is only 600ms behind the best Sweden at the ${75}^{th}$ percentile.

\noindent\textbf{DOM Content Loaded:} Similar to the previous observation, the minimum is again at 0ms for all countries. The ${25}^{th}$ percentile is at 600ms for Germany, Great Britain, France, Sweden and Finland and at 800ms for the other countries. At the median, the results start varying slightly, with Germany, Sweden and Finland having the best results at 1200ms, followed by Great Britain, France, Poland and Romania at 1400ms, and finally Italy and Spain where the median is another 200ms higher. The ${75}^{th}$ percentile has a wider spread with Sweden and Finland having the lowest median at 2200ms. Germany is 200ms higher whereas Great Britain, France, Poland and Romania have a 400ms higher average ${75}^{th}$ percentile than the best ones. The percentile increases again by 200ms for Spain and another 200ms for Italy. The maximum can be grouped in a similar way with Sweden and Finland having a maximum of 4600ms, 
whereas Italy is the worst performing amongst all with 6200ms. All other countries have metrics in between these values.

Even for the metric DCL, Sweden and Finland have the best user experience, closely followed by Germany. 
Again, Italy lacks behind other countries and provides a relatively worse user experience.

\noindent\textbf{Onload:} Again, as for all the other metrics the minimum is at 0ms for every country. The ${25}^{th}$ percentile is below the 1600ms threshold for all countries. The lowest percentile is achieved in Germany, Sweden and Finland with 1000ms, followed by Great Britain, France, Spain and Poland with 1200ms. The remaining countries have a ${25}^{th}$ percentile of 1400ms. Germany, Sweden and Finland had all together the lowest median as well with 2200ms. Great Britain, France, Spain and Poland is next with 2400ms, followed by Romania at 2600ms and Italy at 2800ms. The ${75}^{th}$ percentile is dispersed even further. Its lowest value is 4000ms for Sweden and Finland while highest is recorded for Italy with 5000ms. The other countries rank in the same order as for the median, except Germany which is at 4200ms. The same trend is observed for the maximum. Again, Sweden and Finland have the lowest maximum at 8400ms whereas Italy has the highest maximum at 10400ms. The other countries also maintain their order here.

The metric Onload follows the global trend and QoE order as previously discussed for all the countries. The data shows that OL lacks only a few 100 milliseconds behind the previously analysed metrics, like DCL. This suggests a great improvement in user experience when browsing on a mobile device. The data also suggests that the difference between the metrics will get smaller in the future. Combined with an Onload which gets even smaller, this is a good sign for the overall QoE when measuring loading times. 
  
To summarize, the northern countries (Sweden and Finland) perform well when it comes to user experience. Germany has been able to come close to these countries and increase its mobile user experience, which is great as it has the most number of distinct origins. Countries like Italy improved as well in terms of QoE, but it still has a long road ahead to close the gap with other countries.

\begin{summary}
\takeaway{The country-wise grouping of connection distribution types, shows Sweden topping the list with maximum 4G connection requests (>85\%), followed by Finland ($\approx$81\%) whereas Italy is at the bottom ($\approx$69\%). The temporal analysis at the ${75}^{th}$ percentile shows Sweden and Finland performing 30\% (FCP), 36.36\% (DCL) and 25\% (OL) better than Italy. This reaffirms that Sweden and Finland indeed provide good user experience.
}
\end{summary}

\section{Limitations and Future Work}\label{sec:limitations}

All metrics in the CrUX dataset are aggregated by origin as statistical data in the form of normalized histograms. As a result, it is not discernible whether an origin’s metric quality has increased over time or not. Thus, any improvement on the metrics would purely be of qualitative (i.e. more accurate) nature. There are a few other factors that limit the general validity of these results. Firstly, as the data source for this dataset is the Google Chrome browser which is not the default browser on most available mobile phone OS, users have to actively use/select Google Chrome as their default browser. Secondly, this measurement is only based on a subset of users who particularly did not opt out from sharing their browser history with Google. Lastly, only websites known to the Google web crawler are used as input data by the CrUX report. This may lead to new websites not being considered while older unused websites still providing data.

As a future extension, it would be interesting to explore 
a diverse categorization for connection speeds faster than 700Kbps. It could lead to a more precise analysis of quality of experience (QoE) and thereby to more concise results. Further research can be focused on origin-specific segmentation, as classification of origins can be used for implicit classification of the CrUX data. Going deeper into the origin values, it is possible to measure performance of mobile-specific origins in contrast to their desktop counterparts. A possible follow-up study could also validate the results with a different web browser across all device types. It would be especially interesting for browsers used as default for mobile devices such as Safari. Lastly, studies could investigate factors that explain why the user experience differs between European countries and possibly try to give explanations on which factors to improve on in order to enhance the user QoE.

\section{Related Work}\label{sec:relatedwork}

The understanding of user QoE is vital for the real-world deployment of new products, technologies and processes. The QoE metric that encompasses the objectiveness of the user and methodologies for quantification of human’s perception of the metric have been discussed in \cite{Moorsel}\cite{DBLP:conf/wwic/KuipersKVB10}\cite{DBLP:conf/wwic/Martinez-YelmoSG10}.  The effort to factor in more users, highlighted the need for real world data measurements such as the one conducted by Ickin \emph{et al.} \cite{DBLP:journals/cm/IckinWFJHD12} for mobile application quality of experience. Initially, time-based metrics such as response timings \cite{DBLP:journals/cn/BhattiBK00} or delays were used in QoE measurements, which are dependent on the service layer aspects such as network access quality \cite{DBLP:journals/cm/IckinWFJHD12}\cite{DBLP:journals/isf/ShinLSL10}.

Due to the timing-based QoE metrics, Briscoe \emph{et al.} \cite{DBLP:journals/comsur/BriscoeBPHRTGFG16} surveyed techniques to reduce Internet latency. As real-life active user data measurements often involved high user cost, Ravindranath \emph{et al.} \cite{DBLP:conf/osdi/RavindranathPAMOS12} proposed a utility measuring mobile application critical path during usage, while Aggarwal \emph{et al.} \cite{DBLP:conf/wmcsa/AggarwalHPVY14} proposed a system utilizing passive network measurements to estimate the quality of experience. 

Efforts to improve user experience in mobile Internet browsing regarding loading times and latency have been gathered and reviewed by Cazanas et al. \cite{cazanas2017strategies}, identifying three major web design strategies aimed at heightening mobile user quality of experience: Responsive web design on which work has also been done by Almeida \emph{et al.} \cite{almeida}, adaptive web design and separate site deployment. Novel approaches such as cognitive networks to handle the quality of experience requirements in future Internet evolution have been presented in \cite{DBLP:journals/sj/BruniPKPP16}. Koester \emph{et al.} \cite{DBLP:conf/qomex/KosterMM17} presented and evaluated the quality of experience modeling techniques based on underlying technical quality dimensions.

Very limited work has been done on the Google CrUX dataset till date. In one such work \cite{Ruth_website_IMC'22}, Ruth \emph{et al.} evaluates the relative accuracy of the most popular top lists of websites using a set of popularity metrics derived from server- side requests seen at Cloudflare. They observe that only the CrUX dataset accurately captures the top list of websites compared to Cloudflare across all metrics. Another work \cite{Ruth_WWW_IMC'22}, studies the CrUX dataset to analyze how people spend time on the web. They study the dataset to investigate the distribution of web traffic, the types of websites that people visit often and spend the most time on and also the most popular websites in different regions of the world. However, to the best of our knowledge, there is no
such comprehensive study on users browsing experience across different metrics and dimensions over
the CrUX dataset.

\section{Conclusion}\label{sec:conclusion}

The paper utilized the CrUX dataset to extensively evaluate users' Web browsing experience for four different QoE metrics. Segmenting the dataset by connection type, we quantitatively evaluated users' experience across three types of devices (desktop, phone and tablet) and nine European countries. Our study revealed that desktops significantly outperformed phones, which in turn outperformed tablets across all metrics. Analysis along the country dimension reflected the dominance of North European countries, such as Sweden and Finland over the South European ones like Italy. These two North European countries performed 25\%-36\% better than Italy at the ${75}^{th}$ percentile for all of the popular user experience metrics in the CrUX dataset. This reaffirmed existing observations while strengthening the base for future research.


\bibliographystyle{IEEEtran}
\bibliography{index}






 





\end{document}